\documentclass[12pt]{article}
\usepackage{geometry}
\usepackage{amsmath,amssymb}
\usepackage[nosort]{cite}
\usepackage{float}
\usepackage{latexsym}
\usepackage{graphicx}
\usepackage{color}
\usepackage{subfig}
\usepackage{color}

\newcommand{\beq}{\begin{equation}}
\newcommand{\be}{\begin{equation}}
\newcommand{\ee}{\end{equation}}
\newcommand{\bea}{\begin{eqnarray}}
\newcommand{\eea}{\end{eqnarray}}

\newcommand{\pa}{\partial}

\newcommand{\nn}{\nonumber}

\newcommand{\mc}[1]{{\mathcal #1}}

\newcommand{\Dh}{\hat{D}}
\newcommand{\Qh}{\hat{Q}}

\newcommand{\ma}{\mathcal A}

\begin{document}

\begin{titlepage}
\hbox to \hsize{\hspace*{0 cm}\hbox{\tt }\hss
    \hbox{\small{\tt }}}

\vspace{1 cm}

\centerline{\bf \Large Lifshitz-like black brane}

\vspace{.6cm}

\centerline{\bf \Large  thermodynamics in higher dimensions}

\vspace{1 cm}
 \centerline{\large  $^\dagger\!\!$ Gaetano Bertoldi, $^\dagger\!\!$ Benjamin A. Burrington}
 \centerline{\large $^\dagger\!\!$ Amanda W. Peet and $^\dagger$Ida G. Zadeh}

\vspace{0.5cm}
\centerline{\it ${}^\dagger$Department of Physics,}
\centerline{\it University of Toronto,}
\centerline{\it Toronto, Ontario, Canada M5S 1A7. }

\vspace{0.3 cm}

\begin{abstract}

Gravitational backgrounds in $d+2$ dimensions have been proposed as holographic duals to Lifshitz-like theories describing critical phenomena in $d+1$ dimensions with critical exponent $z\geq 1$.  We numerically explore a dilaton-Einstein-Maxwell model admitting such backgrounds as solutions. Such backgrounds are characterized by a temperature $T$ and chemical potential $\mu$, and we find how to embed these solutions into AdS for a range of values of $z$ and $d$.  We find no thermal instability going from the $T\ll\mu$ to the $T\gg \mu$ regimes, regardless of the dimension, and find that the solutions smoothly interpolate between the Lifshitz-like behaviour and the relativistic AdS-like behaviour.  We exploit some conserved quantities to find a relationship between the energy density $\mc E$, entropy density $s$, and number density $n$, $\mc E=\frac{d}{d+1} \left(Ts+\mu n\right)$, as is required by the isometries of AdS$_{d+2}$.  Finally, in the $T\ll\mu$ regime the entropy density is found to satisfy a power law $s\propto c\, T^{{d}/{z}} \mu^{(z-1){d}/{z}}$, and we numerically explore the dependence of the constant $c$, a measure of the number of degrees of freedom, on $d$ and $z$.

\end{abstract}
\end{titlepage}

\section{Introduction}

By now, holography is considered a general tool for studying many kinds of physical systems.  In extremely symmetric situations, e.g. the AdS$_5\times$ S$^5$ background where AdS/CFT was born \cite{Maldacena:1997re}, the symmetries allow many qualitative and quantitative checks of the duality.  A vast number of setups for superconformal field theories have been studied \cite{Oz:1998hr}, again taking advantage of the symmetries.  Less symmetric situations exist \cite{smallsym}, but these still are designed to address questions relevant for particle physics, and so preserve Poincar\'e symmetry by construction.

Much has been learned from the $3+1$ dimensional cases described above.  It is known that finite temperature in the field theory often corresponds to the presence of a horizon in the gravitational dual.  Further, $U(1)$ gauge symmetries in the bulk correspond to conserved number operators in the dual field theory.  Therefore, to study a field theory at finite temperature and chemical potential in a holographic setup, the focus is on charged objects with horizons in the bulk: charged black holes \cite{bhphases}.

Recently these techniques have been brought to bear on other types of systems.  Most notably, condensed matter systems have been a topic of intense research using holographic techniques \cite{Son:2008ye,Balasubramanian:2008dm,condensedmatter,Volovich:2009yh} (see \cite{Hartnoll:2009sz,Sachdev:2010ch,Herzog:2009xv} for an introduction).  This seems quite natural, given that theoretical condensed matter and particle physics share quantum field theory (QFT) as their primary language.  The intuition gained from the 3+1 dimensional setups suggest that holography should also be useful to study condensed matter systems at strong coupling, further implying that certain condensed matter systems may have regimes of ``stringy'' behaviour.

In particular, much effort has gone into describing quantum critical behaviour of these theories.  As mentioned above, identifying symmetries is of primary importance.  Quantum critical systems typically exhibit a scaling symmetry
\be
t\rightarrow \lambda^{z}t, \quad x_i\rightarrow \lambda x_i
\ee
similar to the scaling invariance of pure AdS ($z=1$) in the Poincar\'e patch.  From a holographic standpoint, this suggests the form of the spacetime metric
\be
ds^2=L^2\left(r^{2z} dt^2+ r^2 dx^i dx^j \delta_{i j}+\frac{dr^2}{r^2}\right), \quad i\in \{1,2...d\} \label{scalemetric}
\ee
where the above scaling is realized as an isometry of the metric along with $r\rightarrow \lambda^{-1} r$.
The above metric has no symmetry that mixes time and space, although there are also models that contain a Galilean (or other non-relativistic) symmetry \cite{Son:2008ye,Balasubramanian:2008dm,Yamada:2008if}, many of which admit (or were constructed using) supergravity embeddings.  It is interesting to note that many of the metrics used for non-relativistic physics are known to be coset spaces \cite{SchaferNameki:2009xr}.

While it may at first appear as though (\ref{scalemetric}) is a simple generalization of AdS, there are some significant differences \cite{Copsey:2010ya}.
Here we will not address such concerns, but it should be noted that the above spacetime has been embedded into 10D and 11D supergravity theories \cite{Blaback:2010pp,Balasubramanian:2010uk,Singh:2010zs,Singh:2010cj,Donos:2010tu,Gregory:2010gx,Donos:2010ax} and it would be interesting to see how, or if, these theories avoid the problems discussed in \cite{Copsey:2010ya}.  Further these models   \cite{Blaback:2010pp,Donos:2010tu,Gregory:2010gx,Donos:2010ax} may shed light onto the weak coupling degrees of freedom, and so show what types of field theories these gravitational systems describe.

Here we will be concerned with phenomenological models that admit the metric (\ref{scalemetric}) as a solution.  Two such models with $d=2$ have Lagrangians given by \cite{Kachru:2008yh}
\be
S'=\frac{1}{16\pi G_4}\int d^4x\sqrt{-g}\left(R-2\Lambda-\frac14 {\mathcal G}^2-\frac{c^2}{2}\ma^2\right)
\ee
(we call this model $S'$ for short) and \cite{Taylor:2008tg}
\be
S=\frac{1}{16 \pi G_4}\int d^4x \sqrt{-g}\left(R-2\Lambda-2(\nabla \phi)^2-e^{2\alpha \phi} {\mathcal G}^2\right),
\ee
(we call this model $S$ for short) where in either case ${\mathcal G}=d \ma$ is a two form field strength.  There are also examples of actions with $R^2$ corrections that admit solutions of the form (\ref{scalemetric}) \cite{Cai:2009ac}.

There are several differences between the above models.  First, the Lifshitz solution to model $S'$ is truly invariant under the Lifshitz rescaling symmetry, while the solution with metric (\ref{scalemetric}) of model $S$ has a logarithmically running dilaton, and so we call this a ``Lifshitz-like'' geometry.  Note also that $S$ has a $U(1)$ gauge symmetry while $S'$ does not, and so the field theory dual to $S$ has a conserved particle number $N$ with chemical potential $\mu$.  Also of note is that model $S$ admits an exact black brane solution \cite{Taylor:2008tg} that asymptotes to the metric (\ref{scalemetric}), which can also be generalized to higher dimensions \cite{Chen:2010kn}.  Finding black brane solutions for action $S'$ has proven more difficult, and one often needs to resort to numeric methods \cite{Danielsson:2008gi,Mann:2009yx,Bertoldi:2009vn,Pang:2009ad}, but not always \cite{Mann:2009yx,Bertoldi:2009dt} (an analogous analytic statement for an $R^2$ extension may be found in \cite{Dehghani:2010gn}, and one should also see \cite{Balasubramanian:2009rx} where a certain extension to this model admits an analytic black hole).  For extensions and variations to the basic model $S'$, see \cite{Dehghani:2010gn,extensions}, and for the holographically renormalized action, see \cite{Ross:2009ar}.

Similar actions to $S$ can be found in \cite{Liu:2010ka,Fadafan:2009an,Sin:2009wi,Cadoni:2009xm,Aprile:2010yb,Charmousis:2010zz, Lee:2010qs, Perlmutter:2010qu,Hartnoll:2007ih,Wen:2010et,Puletti:2010de,Hartnoll:2010ik}, which either contain different matter fields (probe, or back-reacted), different couplings, or some combination.
However, here we will study the action $S$, also studied in \cite{Taylor:2008tg,Pang:2009ad,Goldstein:2009cv,Bertoldi:2010ca}, generalized to arbitrary dimension as in \cite{Chen:2010kn}
\be
S=\frac{1}{16 \pi G_{d+2}}\int d^{d+2}x \sqrt{-g}\left(R-2\Lambda-2(\nabla \phi)^2-e^{2\alpha \phi} {\mathcal G}^2\right).
\ee
We will consider ``UV completing'' by embedding black branes into asymptotically AdS space, generalizing to generic $d$ the discussion of \cite{Bertoldi:2010ca}.  We will explore the thermodynamics of these backgrounds analytically and numerically.

As investigated in our previous work \cite{Bertoldi:2010ca}, there are two scales in the problem, $T$ and $\mu$, and there are two regimes, $T\ll \mu$ (Lifshitz-like) and $T\gg \mu$ (AdS-like).  Therefore, one might anticipate  a thermal instability during the transition between the two regimes.  However, in the $d=2$ case, we found that there was no such instability, and so no discontinuous phase transition.  We thought that this could be related to a Coleman$-$Mermin$-$Wagner theorem applied to the thermal vacua of a $2+1$ dimensional theory.  However, here we find that there is no discontinuous phase transition, and the model smoothly interpolates between the Lifshitz-like behaviour and the AdS-like behaviour, regardless of $d$.

Finally, in the Lifshitz-like regime, we expect a relation of the form
\be
4G_{d+2}s=c(z,d) (L\mu)^d\left(\frac{T}{\mu}\right)^{{d}/{z}}\,,
\ee
where $s$ is the entropy density, $T$ is the temperature, $d$ is the number of spatial dimensions, and $z$ is the critical exponent.  The coefficient $c(z,d)$ is in some sense a measure of the number of degrees of freedom of the system, and we find its behaviour as a function of $d$ and $z$ numerically.

The rest of the paper is organized as follows.  In Section \ref{AnalyticsSection} we reduce the model to an effective radial model for arbitrary dimension $d$, and find that the equations of motion reduce to a set of 4 first order ordinary differential equations.  We comment on the proper normalization of the charge density $q$.  We compute the perturbative expansion at the horizon and at the AdS asymptotic, and use this to show the relation
\begin{equation}
\mathcal{E}=\frac{d}{d+1}\left(Ts+\mu n\right)
\end{equation}
which is necessary by the scaling symmetry of AdS \cite{Bertoldi:2010ca}.

Finally, in Section \ref{discussion}, we turn to numeric results.  We set up the numerical integration, and show how to efficiently parameterize the different regimes of ${T}/{\mu}$ with one piece of data specified at the horizon.
We then numerically integrate, and show that for $d=2,3,4\cdots 9$ that there is no discontinuous phase transition: the solutions smoothly and monotonically interpolate between the two regimes ${T}/{\mu}\ll 1$ and ${T}/{\mu}\gg1$.  We then find $c(z,d)$ for a range of $z$ and $d$, and discuss some of its qualitative behaviour.

\section{Analysis of the model}

\label{AnalyticsSection}

\subsection{Reduction}

We wish to consider dilatonic black brane solutions to the equations of motion following from the action
\be
S=\frac{1}{16 \pi G_{d+2}}\int d^{d+2}x \sqrt{-g}\left(R-2\Lambda-2(\nabla \phi)^2-e^{2\alpha \phi} {\mathcal G}^2\right). \label{baseact}
\ee
We reduce the above action on the following Ansatz
\bea
ds^2=-e^{2A(r)}dt^2+e^{2B(r)}\left(dx^i dx^j\delta_{ij}\right)+e^{2C(r)}dr^2 \nn \\
\phi=\phi(r), \qquad {\mathcal A}=e^{G(r)}dt
\eea
where $i\in \{1,2,\cdots d\}$ and  ${\mathcal G}=d{\mathcal A}$.  We will get only ordinary differential equations in what follows, and so we define $\pa\equiv \frac{\pa}{\pa r}$.  We may reduce the action to a one dimensional action \footnote{We keep track of normalization for later use.}
\be
S=\frac{1}{16 \pi G_{d+2}}\int 2 dt  \prod_{i=1}^d dx_i \int dr L_{1D}
\ee
where the one dimensional Lagrangian is given by
\bea
&&L_{1D}=d e^{A+dB-C}\pa A \pa B+\frac{d(d-1)}{2}e^{A+dB-C}(\pa B)^2+e^{-A+dB-C+2G+2\alpha \phi}(\pa G)^2 \nn\\
&&\qquad \qquad \qquad \qquad \qquad \qquad \qquad \qquad\qquad -e^{A+dB-C}(\pa \phi)^2-e^{A+dB+C}\Lambda. \label{redlag}
\eea
Note that when $d=2$ this Lagrangian agrees with our earlier work.  It can be verified that all equations of motion associated with the action (\ref{baseact}) are reproduced by (\ref{redlag}), as long as one uses the equation of motion for $C$.  Here, $C$ acts as a Lagrange multiplier, imposing the ``zero energy'' condition.  Further, we note that $C$ allows for generic $r$ diffeomorphisms.  We will refer to changing $r$ coordinate as ``coordinate gauge'' transformations, to differentiate from the $U(1)$ gauge transformations associated with ${\mathcal A}$.

There are many conserved quantities associated with the above action.  First, there is the conserved quantity associated with the shift symmetry $(A,B,C,\phi,G)\rightarrow (A+d\delta_1,B-\delta_1,C,\phi,G+d\delta_1)$. This can be understood as a rescaling of the time coordinate, and the $x_i$ coordinates that leaves $dt \prod_i dx_i$ invariant, and this then descends to the 1D action as Noether symmetry.  Further, we note that there is the conserved quantity associated with $e^G\rightarrow e^G+{\rm const}$ which is exactly the global part of the gauge symmetry associated with ${\mathcal A}$.  Further, we note that $(A,B,C,\phi,G)\rightarrow (A,B,C,\phi+\delta_2,G-\alpha\delta_2)$ is also a symmetry.  One can view this as saying that the dilaton here is a space dependent gauge coupling, and we may absorb a normalization of the gauge coupling into the definition of ${\mathcal A}$.  In addition, we have the Hamiltonian constraint as always.  Now we count $4$ dynamical fields, and $4$ conserved quantities.  This implies that the system is completely equivalent to a set of first order differential equations
\bea
&&  d e^{A+dB-C}\pa A \pa B+\frac{d(d-1)}{2}e^{A+dB-C}(\pa B)^2+e^{-A+dB-C+2G+2\alpha \phi}(\pa G)^2 \nn\\
&&\qquad \qquad \qquad \qquad \qquad \qquad-e^{A+dB-C}(\pa \phi)^2+e^{A+dB+C}\Lambda=0, \\
&& d e^{A+dB-C}\pa A-d e^{A+dB-C}\pa B-2 d e^{-A+dB-C+2G+2\alpha\phi}\pa G= {\mathcal{D}}_0, \\
&&2 e^{A+dB-C}\pa \phi+2 \alpha e^{-A+dB-C+2G+2\alpha\phi}\pa G={\mathcal{P}}_0, \\
&&e^{-A+dB-C+G+2\alpha\phi}\pa G= Q.
\eea
We form the combination $D_0=\alpha {\mc D}_0+d{\mc P}_0, P_0=\frac 12 {\mc P}_0$ and find
\bea
&&  d e^{A+dB-C}\pa A \pa B+\frac{d(d-1)}{2}e^{A+dB-C}(\pa B)^2+e^{-A+dB-C+2G+2\alpha \phi}(\pa G)^2 \nn\\
&&\qquad \qquad \qquad \qquad \qquad \qquad-e^{A+dB-C}(\pa \phi)^2+e^{A+dB+C}\Lambda=0, \label{eom1}\\
&& d\left(2 \pa \phi+ \alpha\left(\pa A- \pa B\right)\right) e^{A+dB-C}= D_0, 
\label{eom2}\\
&&e^{A+dB-C}\pa \phi+\alpha e^{-A+dB-C+2G+2\alpha\phi}\pa G=P_0, \label{eom3}\\
&&e^{-A+dB-C+G+2\alpha\phi}\pa G= Q. \label{eom4}
\eea
There are two known one parameter families of solutions.  First, there is the AdS black brane
\bea
&&A(r)=\ln\left(L r \sqrt{1-\left(\frac{r_h}{r}\right)^{d+1}}\right), \qquad B(r)=\ln(Lr), \nn \\ && C(r)=\ln\left(\frac{L}{r\sqrt{1-\left(\frac{r_h}{r}\right)^{d+1}}}\right), \nn \\
&& \phi(r)=\phi_b,\qquad  {\mathcal A}=g_b dt.
\eea
where $\Lambda=-{d(d-1)}/({2L^2})$, and $g_b$ and $\phi_b$ are arbitrary constants.  For this background, the conserved quantities are $Q=0$, $D_0= {d(d+1)L^d \alpha r_h^{d+1}}/2$ and $P_0=0$.  Further, the solution has $\hat{T}=LT={r_h(d+1)}/{(4\pi)}$ by reading off the periodicity of imaginary time at the horizon.  Further, using the area law for the entropy density, we find that $s={r_h^d}/{(4 G_{d+2})}$, or in thermodynamic terms $s={(4\pi)^{d}T^dL^d}/{\left[(d+1)^d{4 G_{d+2}}\right]}$.  This shows that the number of degrees of freedom in the field theory is linear in ${L^d}/{(4 G_{d+2})}$.

There is also the Lifshitz black brane given by \cite{Taylor:2008tg,Chen:2010kn}
\bea
&& A(r)=\ln\left(L_L a_L r^z\sqrt{1-\left(\frac{r_h}{r}\right)^{z+d}}\right),\qquad  B(r)=\ln(Lr), \nn \\
&& C(r)=\ln\left(\frac{L_L}{r\sqrt{1-\left(\frac{r_h}{r}\right)^{z+d}}}\right), \qquad 2 \alpha \phi(r)=\ln\left(r^{-2d}\Phi\right), \nn \\
&& G(r)=\ln\left(\frac{(z-1)L^d a_L r^{z+d}\left(1-\left(\frac{r_h}{r}\right)^{d+z}\right)}{2Q}\right) .
\eea
The constants are given by
\be
L_L^2=L^2\frac{(z+d)(z+d-1)}{d(d+1)}, \qquad \Phi=\frac{2\left({Q}/{L^{d-1}}\right)^2(z+d-1)}{d(d+1)(z-1)},
\ee
$a_L$ is arbitrary (and may be removed by time rescaling), and $\alpha=\sqrt{{2d}/{(z-1)}}$.
The conserved quantities are $P_0=0$,
$D_0=\sqrt{{d}/\left[{2(z-1)}\right]}r_h^{z+d}a_L L^d d(z+d)$ and $Q$ (which is arbitrary).
Given that $B(r)$ is normalized in the same way as the AdS black brane, we read $s={r_h^d}/{[4 G_{d+2}]}$.  Finally, we see that $\hat{T}={r_h^z a_L (z+d)}/{(4\pi)}$.  While this is a dimension free measure of temperature, it is not clear what units to use.  This is to be expected for a non-relativistic theory because energies and length scales are not interchangeable.  However, this ambiguity is only a constant, and so $s\propto T^{{d}/{z}}$.  When we consider embedding these black branes into AdS space the ambiguity is removed.

\subsection{Perturbation theory at the horizon}

We first expand the functions as ``correction functions'' that should go to constant values for an AdS background.
We therefore define
\bea
&& A(r)=\ln(Lr)+A_1(r), \quad B(r)=\ln(Lr),\nn \\
&& C(r)=\ln\left(\frac{L}{r}\right)+ C_1(r), \quad G(r)=\ln(L)+ G_1(r).
\eea
We further define dimensionless conserved quantities
\be\label{consquant}
Q=L^{d-1}\Qh, \qquad D_0=L^d \Dh_0, \qquad P_0=L^d \hat{P}_0=0
\ee
where we remind the reader that we will only use the $P_0=0$ gauge.  We have 4 first order differential equations, but we only have 3 dynamical functions.  Therefore, we may eliminate one function using an algebraic expression, and we do so for the field $e^{-2\alpha \phi}$:
\bea\label{phiexplode}
&& e^{-2\alpha \phi}=\frac{1}{2 r^2 \Qh^2 \alpha e^{A_1+C_1}} \Bigg(d r^{2d+2}\alpha(d+1)\left(e^{A_1+C_1}-e^{A_1-C_1}\right) \nn \\
&& \qquad \qquad \qquad -2r^{d+1}\Dh_0-4dr^{d+1}e^{G_1}\alpha \Qh+2\alpha e^{C_1-A_1}(\alpha \Qh e^{G_1})^2\Bigg).
\eea
Using this, the other differential equations may be written
\bea
&& \pa e^{A_1} -\frac{e^{C_1}\left(\Dh_0 + 2d\alpha \Qh e^{G_1}\right)}{d\alpha r^{d+2}}=0,  \label{A1numint}\\
&& \pa e^{C_1} -\frac{e^{2C_1-2A_1}\left(2\alpha e^{C_1}\alpha^2 \Qh^2 e^{2G_1}-2 d e^{A_1}r^{d+1}\alpha \Qh e^{G_1} - \Dh_0 e^{A_1}r^{d+1}\right)}{d \alpha r^{2d+3}}=0, \label{C1numint}\\
&& \pa e^{G_1}-\frac{\alpha d r^{2d+2}}{2\alpha \Qh r^{d+2}}\Bigg((d+1)\left(e^{A_1+C_1}-e^{A_1-C_1}\right) \label{G1numint} \\
&& \qquad \qquad \qquad -2r^{d+1}\left(\Dh_0+2d\alpha\Qh e^{G_1}\right)+2\alpha e^{C_1-A_1}\alpha^2 \Qh^2 e^{2G_1}\Bigg)=0. \nn
\eea
These functions may be expanded around a regular horizon as
\bea
e^{A_1}=a_0\left((r-r_h)^{\frac12}+a_1(r-r_h)^{\frac32}+\cdots\right),\nn \\
e^{C_1}=\frac{c_0}{(r-r_h)^{\frac12}}+c_1(r-r_h)^{\frac12} \cdots ,\nn \\
e^{G_1}=a_0\left(g_0(r-r_h)+g_1(r-r_h)^2+\cdots\right).
\eea
One may read that the dilaton goes to a constant.  With $\hat{P}_0=0$, we find that the equations of motion require
\bea
&& a_0=\frac{2c_0 \Dh_0}{\alpha d r_h^{2+d} },\quad a_1=\frac{\alpha^2d(d+1)^2c_0^4-2dr_h(\alpha^2-2)(d+1)c_0^2+r_h^2\left(\alpha^2 d -4 -6d\right)}{8r_h^3}, \nn \\
&& c_0=c_0, \;\;  c_1=\frac{c_0\left(3\alpha^2d(d+1)^2c_0^4-2dr_h(2+3\alpha^2)(d+1)c_0^2+r_h^2\left(3\alpha^2 d +4+6d\right)\right)}{8r_h^3}, \nn \\
&& g_0 = \frac{d((d+1)c_0^2-r_h)r_h^d}{2c_0 \Qh},  \\
&& g_1=\frac{d^2}{8\Qh c_0r_h}\Bigg(r_h^{d-2}\alpha^2(d+1)^3c_0^6 -r_h^{d-1}\alpha^2(d+1)^2c_0^4 \nn \\
&& \qquad \qquad \qquad \qquad -r_h^{d}(\alpha^2+2)(d+1)c_0^2+r_h^{d+1}\left(\alpha^2+2\right)\Bigg).\nn
\eea
This will provide initial conditions for the equations of motion when we numerically integrate.  From this expansion we can read the temperature and entropy density
\be\label{tempentr}
T=\frac{r_h^2 a_0}{4\pi L c_0}=\frac{\Dh_0}{\alpha d r_h^{d}2 \pi L},\qquad s=\frac{r_h^d}{4 G_{d+2}}.
\ee
Further, we find using the above type of expansion that the Lifshitz black brane satisfies $c_0=\sqrt{\alpha^2+2}\sqrt{{r_h}/{(d+1)}}$ while the AdS black brane satisfies $c_0=\sqrt{{r_h}/{(d+1)}}$.  We expect that these two limits bound the physical values of $c_0$.

\subsection{Perturbation theory at $r=\infty$: AdS asymptotics}
We want to expand the solution about infinity and require that it asymptotes to the pure AdS solution. For this, we use the expansion:
\begin{eqnarray} \label{eq:ptads}
A(r)&=&\ln (Lr)+A_1(r), \nonumber \\
B(r) &=& \ln(Lr), \nonumber \\
C(r) &=& \ln \left( \frac{L}{r} \right) +C_1(r), \\
\phi(r) &=& \ln(\Phi_b)+\phi_1(r), \nonumber \\
G(r) &=& \ln(g_b)+G_1(r) \nonumber,
\end{eqnarray}
where now $A_1(r)$, $C_1(r)$, $G_1(r)$, and $\phi_1(r)$ are understood to be {\em perturbative} functions.
Inserting equations (\ref{eq:ptads}) in the equations of motion 
(\ref{eom1})-(\ref{eom4}) and performing the
integrations gives (to leading order)
\begin{eqnarray}
A_1(r)&=&-\frac{D_0+2d\alpha Qg_b-2dP_0}{d(d+1)\alpha L^dr^{d+1}}+
\frac{Q^2}{d(d-1)\Phi_b^{2\alpha}L^{2(d-1)}r^{2d}}, \label{eq:a1}\\
C_1(r)&=&\frac{D_0+2d\alpha Qg_b-2dP_0}{d(d+1)\alpha L^dr^{d+1}}-
\frac{2Q^2}{(d-1)(d+1)\Phi_b^{2\alpha}L^{2(d-1)}r^{2d}},  \label{eq:c1}\\
G_1(r)&=&-\frac{Q}{(d-1)g_b\Phi_b^{2\alpha}L^{d-2}r^{d-1}}, \label{eq:g1}\\
\phi_1(r)&=&\frac{\alpha Qg_b-P_0}{(d+1)L^dr^{d+1}}-\frac{\alpha Q^2}{2d(d-1)
\Phi_b^{2\alpha}L^{2(d-1)}r^{2d}} \label{eq:phi1}.
\end{eqnarray}
The constants of integration of equations (\ref{eom3}) and (\ref{eom4}) (which would lead to constant modes in (\ref{eq:phi1}) and (\ref{eq:g1}) ) are absorbed in the boundary values of $\Phi_b$ and $g_b$ , respectively.
The constant of integration of equation (\ref{eom2}) is removed by rescaling the time coordinate.

We next want to evaluate the energy density.
We follow closely our previous discussion for the $d=2$ case \cite{Bertoldi:2009vn}, and generalize it here to arbitrary $d$.
We use the background subtraction technique in \cite{Hawking:1995fd}, where the
total energy density is given by
\begin{equation}\label{eq:eI}
\mathcal{E}=-\frac{1}{8\pi}\left(N_t\left(^dK-^dK_0\right)-N_t^{\mu}p_{\mu\nu}\hat{r}^{\nu}\right),
\end{equation}
where $N_t$ is the lapse function, $N_t^{\mu}$ is the shift vector, $^dK$ is the extrinsic
curvature of the $d$-dimensional spatial boundary slice inside the constant-$t$ slice,
$^dK_0$ is the $d$-dimensional extrinsic curvature of the reference background,
$p_{\mu\nu}$ is the momentum conjugate to the time derivative of the metric on the constact-$t$
slice, and $\hat{r}$ is the spatial unit vector normal to the constant-$r$ surface.
The reference background  we take is pure AdS in the Poincar\'e patch.

For our metric, $N_t=e^A$, $N_t^{\mu}=0$, $\hat{r}=e^C$, and the energy density is
\begin{equation}\label{eq:eII}
\mathcal{E}=\mathrm{lim}~~-\frac{1}{8\pi}e^A\left(\frac{d}{2}~\partial_r\left(e^{2B}\right)e^{(d-2)B}
\left( e^{-C} - e^{-C_{ref}}\right) \right)\mid_{r\to \infty}.
\end{equation}
For the pure AdS reference, we have $e^{C_{ref}}=\frac{L}{r}$.
The total energy density then reads
\begin{equation}\label{eq:eIII}
\mathcal{E}= \lim_{r\rightarrow \infty} \frac{2d~r^{d+1}}{16\pi G_{d+2}L}C_1\,,
\end{equation}
where we have dropped terms that go to zero in the large $r$ limit. Therefore, we read the energy density to be
\begin{equation}
\mathcal{E}=\frac{1}{16\pi G_{d+2}}\frac{2}{(d+1)}\frac{(D_0+2d\alpha Q g_b)}{\alpha L^{d+1}}. \label{eq:eIV}
\end{equation}

Now we need to normalize the charge density $Q$.
For this, we follow the discussion of \cite{Batrachenko:2004fd},
where they use the action to normalize $Q$ in such a way that it is conjugate to the fixed field value $e^G$ at infinity.
In our context, this reads
\begin{equation}
S_{\mathcal{G}^2}=\frac{V_d}{2}\frac{\mu_{geom}}{T}q_{geom},\label{eq:sf2can}
\end{equation}
where $q_{geom}$ is a normalized charge density and $\mu_{geom}$ is the potential difference between the horizon and the boundary.  For us, we are using a gauge where $P_0=0$ and so the potential at the horizon is $0$, and so $g_b=\mu_{geom}$ for this gauge choice.  We calculate $S_{\mathcal{G}^2}$ and find
\begin{equation}
S_{\mathcal{G}^2}=\frac{1}{16\pi G_{d+2}}\frac{2V_d}{L^{d+1}}\frac{\mu_{geom}}{T}Q,\label{eq:sf2II}
\end{equation}
and so we identify the charge $Q$ as
\begin{equation}
Q=\frac{16\pi G_{d+2} L^{d+1}}{4}q_{geom}. \label{eq:qnorm}
\end{equation}
Inserting the scaled quantities (\ref{consquant}), and using the relations  (\ref{tempentr}) for the temperature and entropy density along with the normalization of $Q$, we find that 
\begin{equation}
\mathcal{E}=\frac{d}{d+1}\left(Ts+\mu n\right),
\end{equation}
where we have changed to field theory quantities $\mu_{geom}=\mu L^2, q_{geom}={n}/{L^2}$.  This is the expected relation for a conformal field theory.

\subsection{Other considerations, and setup for numerics}

Given the discussion of perturbation theory around the horizon, we find that the constant $c_0$ has the following window of allowed values
\be
\sqrt{\frac{r_h}{d+1}}<c_0< \sqrt{(2+\alpha^2)}\sqrt{\frac{r_h}{d+1}}\,, \label{c0range}
\ee
where the lower limit gives the pure AdS black brane, and the upper limit gives the pure Lifshitz black brane condition.
Using this, we may efficiently parameterize all black branes in terms of this one quantity, which we now explain.

First, we may rescale $r$ by a constant: this will affect the location of the horizon, and effectively allows us to fix the horizon to be at $r_h=1$.  After doing so, we may use time rescaling to set $D_0$ to be any value we wish, and further we may use the global symmetry associated with shifting the dilaton to move $Q$ to be any value we wish.  Therefore, we may fix all of the horizon data except for $c_0$ using symmetries.  Then, $c_0$ parameterizes the different black branes, and the allowed region of $c_0$ is given by the above range (\ref{c0range}).  However, now the asymptotic values of the fields are non-canonical.  This means that, given a $c_0$, we take these asymptotic values of the fields $e^{A}, e^{G}, e^\phi$ to be the output values of the numeric integration started at the horizon.  These output values encode how to use the time rescaling and the global symmetry to bring them to their canonical values.  Hence, these parameterize a given $T$ and $\mu$ for a fixed $r_h=1$ black brane.  Near the limiting values of the range  (\ref{c0range}) we expect the ratio ${T}/{\mu}$ to go to zero (Lifshitz-like regime) or to infinity (AdS regime), and so all possible ratios of ${T}/{\mu}$ are explored.

To find the generic black brane with temperature $T$ and chemical potential $\mu$, we would simply find the appropriate black brane at $r_h=1$ with the same value of ${T}/{\mu}$. We would then use the scaling symmetry of AdS to adjust, say $T$, to its correct value.  The resulting black brane has the specified value of $T$ as its temperature, $\mu$ as its chemical potential, and the position of the horizon $r_h$ will be determined by the rescaling.  These will all be unique as long as the ratio ${T}/{\mu}$ occurs only once for the $r_h=1$ reference black brane.
The crucial question is then whether the graph of ${T}/{\mu}$ vs. $c_0$ is monotonic for the $r_h=1$ black branes. If it is, there is always a unique black brane given a particular value of $T$ and $\mu$, and so there are no possible discontinuous phase transitions.  Indeed, we find this is the case.
For the sake of clarity, instead of graphing the quantity  ${T}/{\mu}$ as a function of the somewhat esoteric quantity $c_0$, we simply graph $s$ vs. $T$ for fixed $\mu$. We will see that it is monotonic, and interpolates between the   Lifshitz-like scaling $s\propto T^{{d}/{z}}$ and the AdS behaviour $s\propto T^{d}$. Accordingly, there is no discontinuous phase transition.

Next, we recall a subtlety that emerged in \cite{Bertoldi:2010ca} associated with the definition of the chemical potential $\mu$, and we recall the resolution here.  First, the physical value of $e^{G_1}$ is not strictly determined.  We may use the global symmetry $(A,B,C,\phi,G)\rightarrow (A,B,C,\phi+\delta_2,G-\alpha\delta_2)$ to rescale this value to any value we wish.  Because this is a global symmetry that does not involve the metric, the stress energy tensor (and therefore the geometry), is not determined by this number.  In fact, only global symmetry invariants can determine anything in the geometry (this was also noticed in \cite{Goldstein:2009cv} for extremal solutions).  We wish to consider $\mu$ as a scale in the theory, i.e. it is the scale at which new particles can be added/excited, and so we expect this to correspond to some scale in AdS, and so must be a global symmetry invariant.

One way to fix this ambiguity is to note that the gauge coupling at infinity is the asymptotic value of $e^{2 \alpha \phi}$ (which also transforms under the global symmetry).  Hence, we would like the gauge kinetic term to go to a canonical value.  This can be arranged by using the global symmetry to scale $e^{2 \alpha \phi}|_{r=\infty}=1$.  From the beginning we should have expected such a statement: only fixing both the asymptotic value of $e^{\alpha \phi}$ {\it and} the asymptotic value of $e^{G_1}$ will determine the geometry.  We had traded this for $e^{G_1}$ and $\hat{Q}$ in the above discussions.  Therefore, one can read the output value of $\mu$ from the numeric integration as
\be
\hat{\mu}= \frac{e^{G_1} e^{\alpha \phi}}{e^{A_1}}\Bigg|_{r=\infty}.
\ee
The factor of $e^{A_1}$ in the denominator accounts for the fact that $e^{G_1}$ transforms under the time rescaling used to set the asymptotic value of $e^{A_1}$ to be $1$, and the factor of $e^{\alpha \phi}$ in the numerator is there because of using the global symmetry to set $e^{\alpha \phi}=1$.  Let us assume that $e^{A_1}$ goes to 1 at infinity.  Then the above simply states that what we have done is take some bare quantities $q_{\rm geom}$ and $\mu_{\rm geom}$ and combined them into the global symmetry invariant $q_{\rm geom}\mu_{\rm geom}= \left(e^{\alpha \phi}|_{r=\infty}q_{\rm geom}\right)\left(e^{-\alpha \phi}|_{r=\infty}\mu_{\rm geom}\right)$.  The second expression is made out of global symmetry invariants (e.g $\left(e^{\alpha \phi}|_{r=\infty}q_{\rm geom}\right)$), and so can define scales in the geometry.  These are what we use to define the physical chemical potential $\mu_{\rm geom}$, and why we will find $e^{\alpha \phi}|_{r=\infty}=1$ a convenient representative from all solutions related by the global symmetry.

With the above considerations, and the perturbative results near the horizon, we may use the equations (\ref{A1numint})-(\ref{G1numint}) and numerically integrate them from the horizon to a large value of $r$, where the functions settle to their AdS asymptotic values.  These values at infinity furnish the required information to determine $T$ and $\mu$ for the black brane.


\section{Results and Discussion}


\label{discussion}

\begin{figure}[ht!]
\centering
\subfloat[$d=3$.]{\label{fig:muh1d3}
\includegraphics[width=0.45\textwidth]{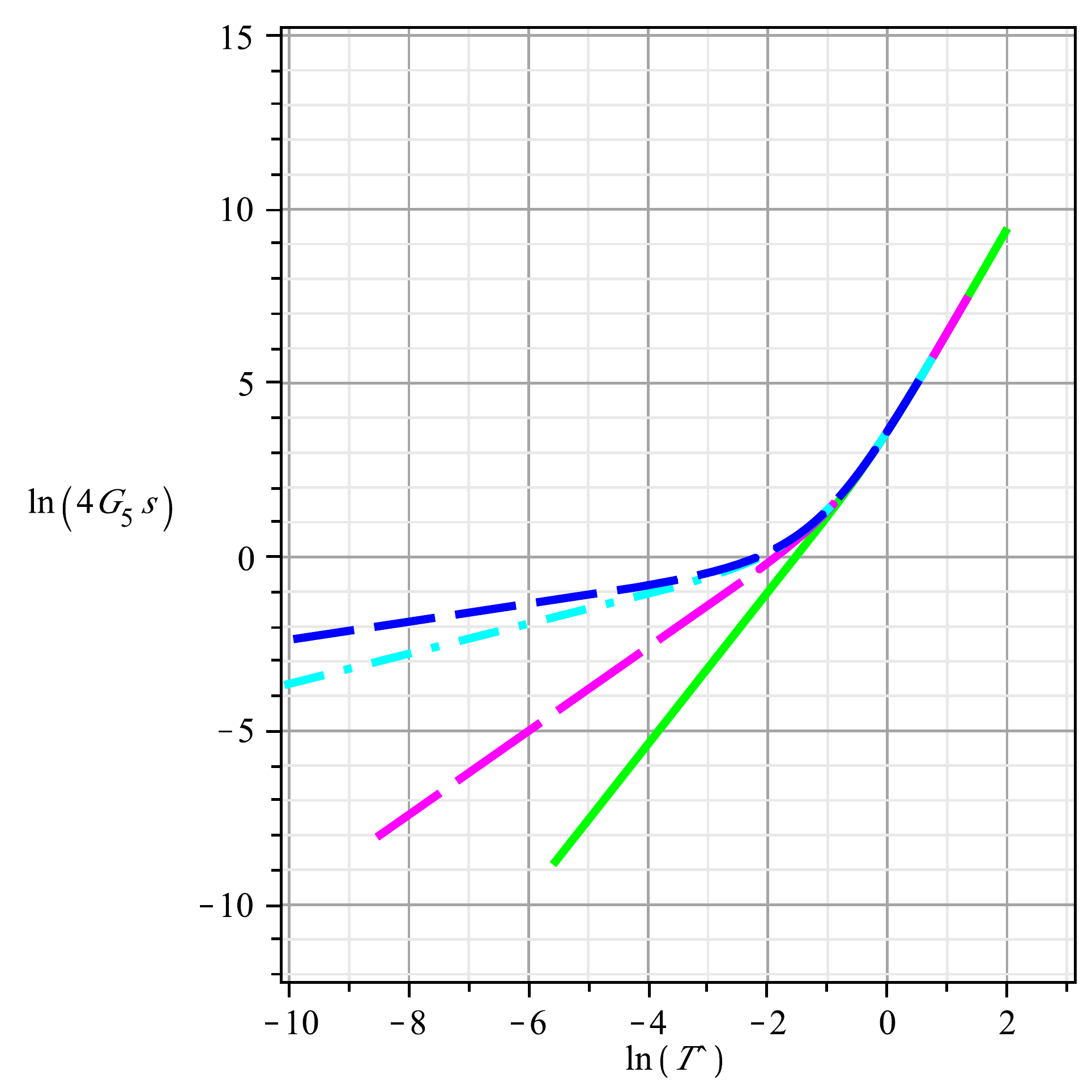}}
\subfloat[$d=5$.]{\label{fig:muh1d5}
\includegraphics[width=0.45\textwidth]{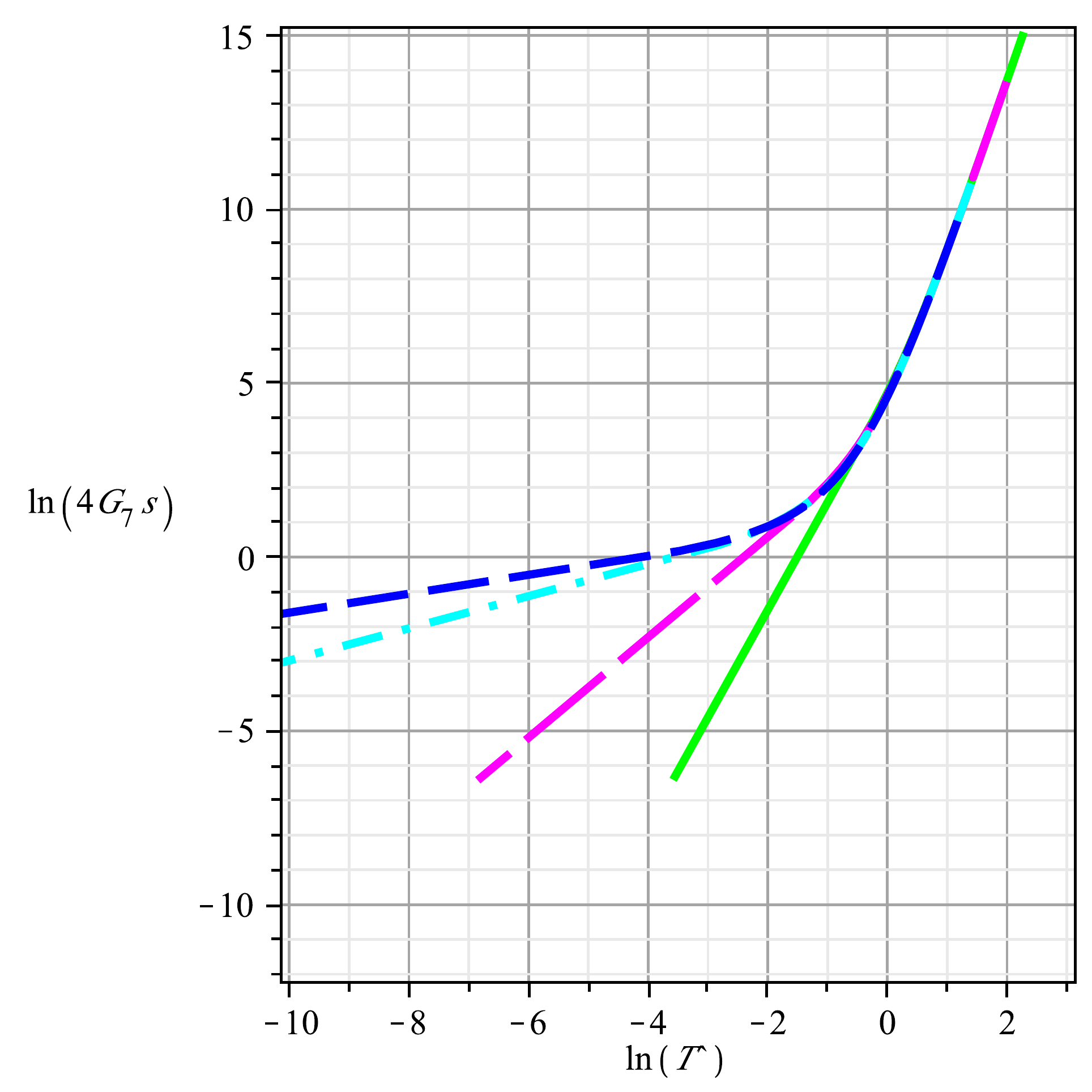}}\\
\vspace{0.1in}
\subfloat[$d=7$.]{\label{fig:muh1d7}
\includegraphics[width=0.45\textwidth]{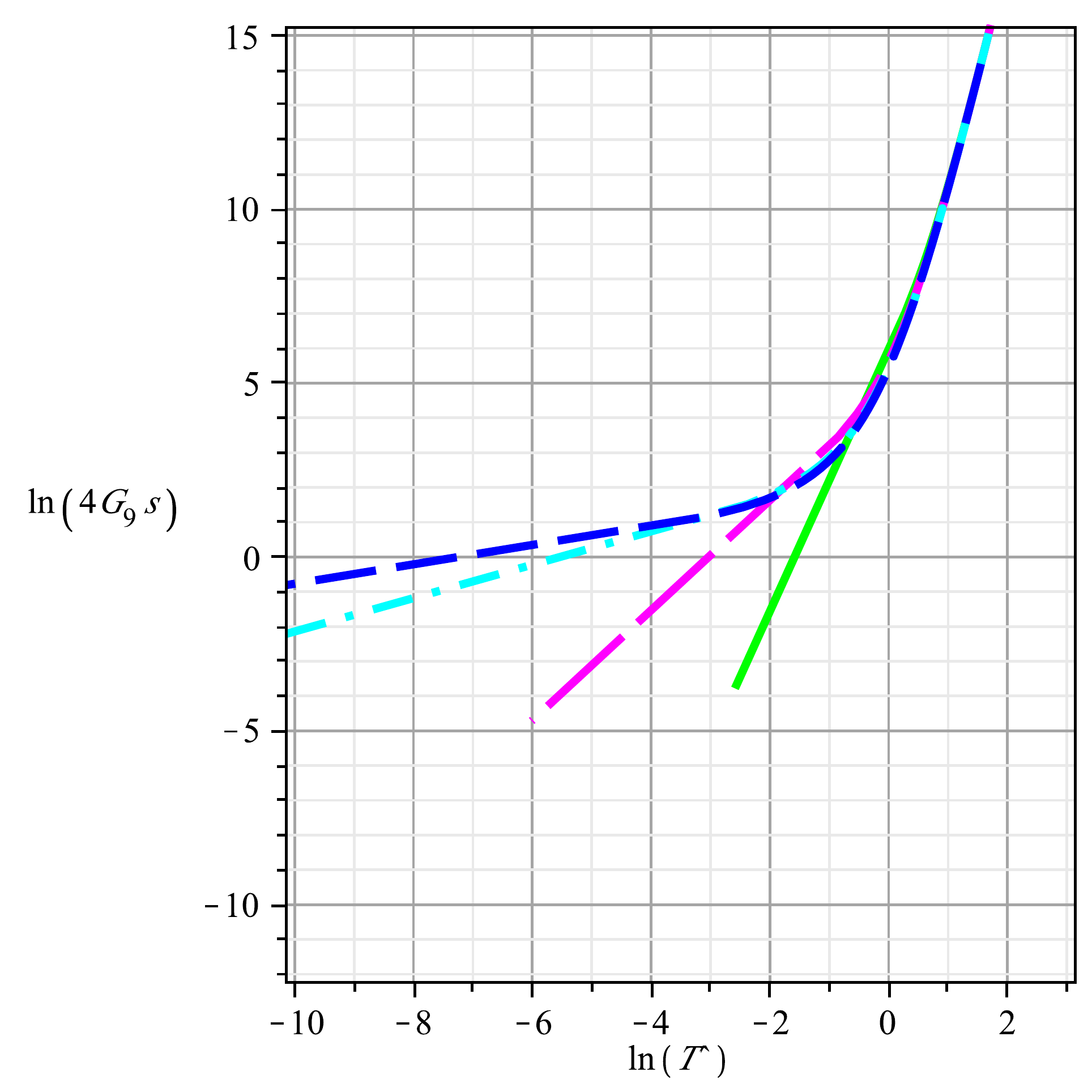}}
\subfloat[$d=9$.]{\label{fig:muh1d9}
\includegraphics[width=0.45\textwidth]{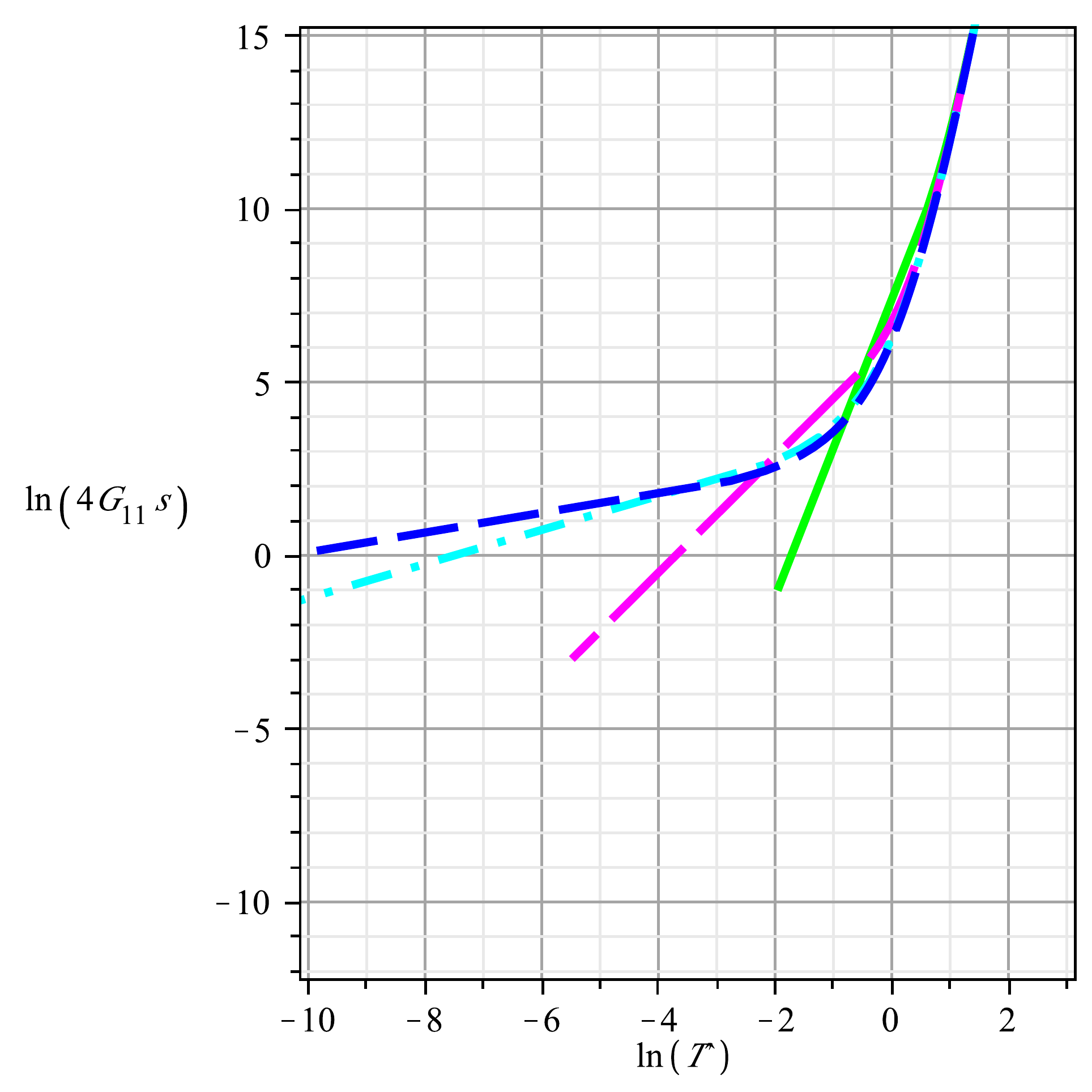}}
\caption{The plots of $\ln\left( 4G_{d+2}{s}\right)$ versus $\ln ({LT})$ for fixed
$\hat{\mu}=1$.
Figures (\ref{fig:muh1d3}), (\ref{fig:muh1d5}), (\ref{fig:muh1d7}) and (\ref{fig:muh1d9})
correspond to $d = 3$, $5$, $7$, and $9$, respectively.
The different curves in each plot correspond to different values of $\alpha$,
with $\alpha=4$ green (solid),
$\alpha=2$ magenta (long-dashed),
$\alpha=1$ cyan (dot-dashed), and
$\alpha=0.75$ blue (dashed).
$\alpha$ is related to $z$ via $\alpha=\sqrt{2d/(z-1)}$.}
\label{fig:s_T}
\end{figure}

Here we discuss the results found from numerically integrating the equations of motion.  In Figure \ref{fig:s_T}, we give log-log plots of the dimensionless entropy density ($4G_{d+2}s$) versus the dimensionless temperature $\hat{T}$ for fixed $\hat{\mu}=1$ and various dimensions.
As seen in Figure \ref{fig:s_T}, the entropy density is smooth and
monotonic in $\hat{T}$ for a wide range of $z$ and there are no possible discontinuous phase transitions
associated with going from $\hat T\ll \hat \mu$ to $\hat T \gg \hat \mu$.
The same behaviour is observed for the other dimensions $2\leq d \leq 9$ not plotted in Figure \ref{fig:s_T}.
Furthermore, we observe the correct asymptotic behaviour: the slope
approaches $d/z$ in the Lifshitz-like regime ($\hat T \ll \hat \mu$); and the slope approaches $d$ in the AdS regime ($\hat T \gg \hat \mu$).

We had thought from our earlier work \cite{Bertoldi:2010ca} that the behaviour in $d=2$ might follow from the relatively low dimension of the dual model.  However, the results of Figure \ref{fig:s_T} show  that this is not the case.
We see no evidence for a discontinuous phase transition or thermal instability regardless of the dimension.
%
To be sure of these results required careful numerical examination of the dilaton  at infinity to ensure that it was indeed smooth and approaching a constant value, as is appropriate for AdS. This is because the expression (\ref{phiexplode}) for $\phi$  has explicit $r$ dependence, and so rounding errors will eventually occur for sufficiently large $r$. The powers of $r$ involved are dimension dependent, so the rounding effect is more severe the higher  the dimension.

\begin{figure}[ht!]
\centering
\includegraphics[width=0.65\textwidth]{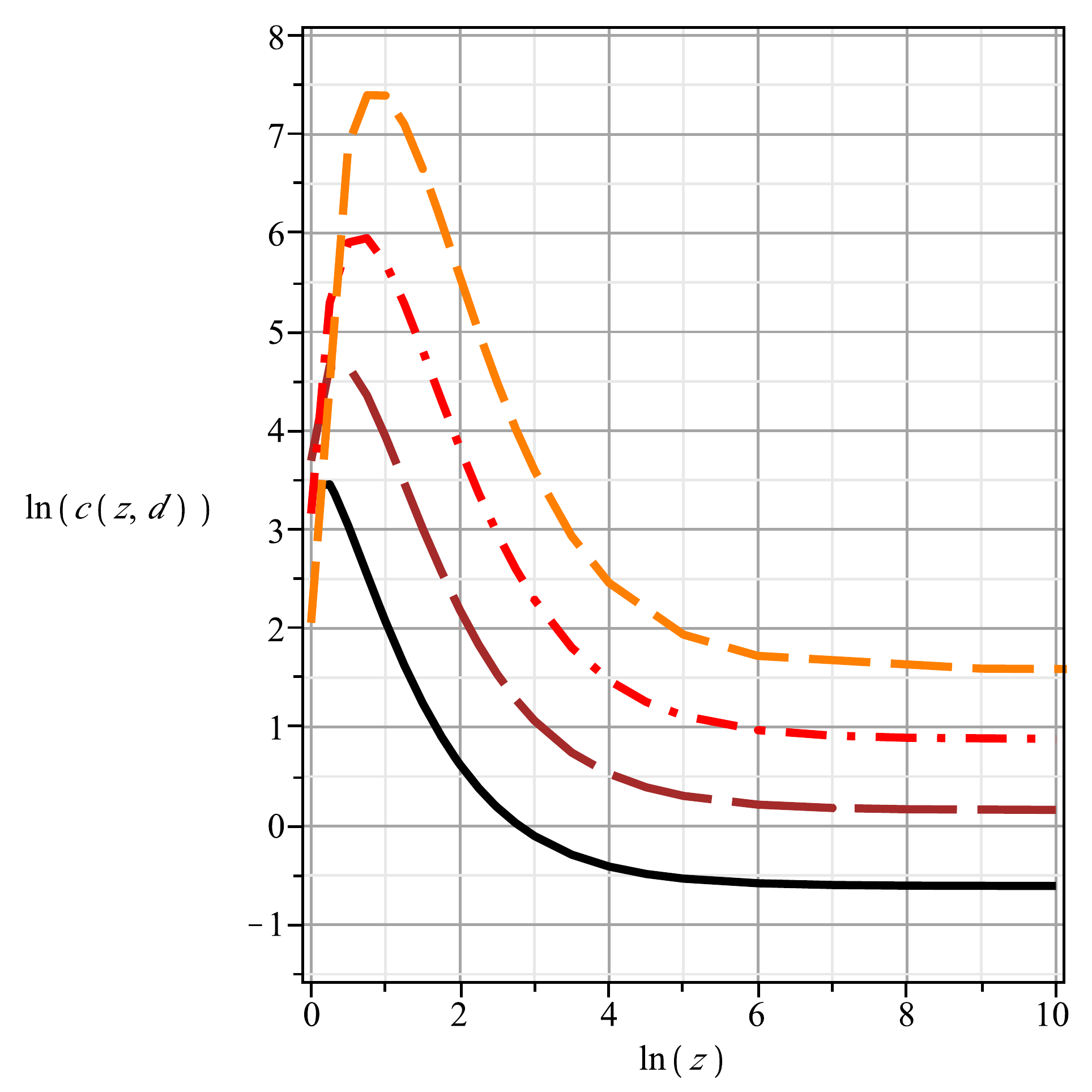}
\caption{The plot of $\ln\left( c(z,d)\right)$ as a function of $\ln (z)$
for fixed value of $\hat{\mu}=1$.
The different curves correspond to different dimension with $d=3$ black (solid), $d=5$ brown (long-dash), $d=7$ red (dot-dash), and $d=9$ coral (dash).
Curves for $d=2,4,6,8$ behave similarly.
First of all, notice that the curves' intercept at $\ln(z)=0$ is given by the value
$
\ln\left({(4\pi)^d}/{(d+1)^d}\right)
$
which is non-monotonic in $d$.  Secondly, we find it interesting that the tails become flat out at $z\rightarrow\infty$ and that
the large $z$ behaviour for various $d$ is monotonic in $d$.  
Given these two facts, it may be no surprise that the curves $\ln(c(z,d))$ 
for fixed $d$ are generically non-monotonic for small values of $z$, and that in fact they cross each other.
}

\label{fig:int}
\end{figure}

Next we turn to the question of measuring the number of degrees of freedom for the Lifshitz-like theories.  We note first that the entropy density of the system in the Lifshitz-like limit is of the form
\begin{equation}\label{eq:s}
4 G_{d+2}s=c(z,d)(L\mu)^{d}\left(\frac{T}{\mu}\right)^{{d}/{z}},
\end{equation}
where $c(z,d)$ is a function of the dimension and the critical exponent.
We have chosen $\hat{\mu}=1$ in the plots shown in Figure \ref{fig:s_T}, and so the above expression reduces to $4 G_{d+2}s=c(z,d)\,{\hat T}^{{d}/{z}}$.  Therefore, if we fit the $\ln(\hat{T})\ll \ln(\hat{\mu})$ portion of the graphs in Figure \ref{fig:s_T} with a line,  the intercept value of this line gives $\ln(c(z,d))$.  This coefficient  is a direct measure of the degrees of freedom of the theory.
Figure \ref{fig:int} depicts the behaviour of $\ln(c(z,d))$ versus $\ln(z)$ for various
dimensions.

We can get a qualitative picture of how these must behave in the large $z$ limit.  First, note that as $z$ becomes large, the slope of the $\hat T \ll \hat \mu$ part of the graph, $d/z$, goes to zero.  The intercept, therefore, approaches a fixed value, and so $c(z,d)$ must asymptote to a constant number.  This is a trend we do observe in  Figure \ref{fig:int}.  Therefore, $c(z,d)$ only depends on the dimension of the theory at large $z$.  

We cannot compare the values of $c(z,d)$ we found for different $z$ and different $d$, since they correspond to different theories. 
Nonetheless, it would be interesting to compare the values of $c(z,d)$ we have found here for model $S$ with the same quantities in different models. 
We leave this to future work.

In the following, we will compare the large $z$ behaviour of the graphs of Figure \ref{fig:int} with expectations from the dual field theory, using a simplified model.  We start by considering a $(2\pi \ell)^d$ volume box of some medium, and consider the excitations above the ground state.  For such a case, the ``legal'' wave numbers may be written
\be
\vec{k}=\begin{pmatrix}
{n_1}/{\ell} \\
{n_2}/{\ell} \\
\vdots
\end{pmatrix}.
\ee
Further we assume Lifshitz scaling symmetry and rotational symmetry in the spatial dimensions (broken only by the presence of the ``box'').  The only consistent relation between the energy of a state and the wave numbers above is
\be
\omega_{\vec{n}}=(\sqrt{k^2})^{z}\mu^{1-z}\delta=(\sqrt{n^2})^{z}\mu \frac{1}{(\ell \mu)^z}\delta \,,
\ee
where the powers of $\mu$
(an intensive parameter) appear by dimensional analysis.  The coefficient $\delta$ will in general be a function of $z$ and $d$; for the time being we will leave this implicit, because we expect that $\delta$ should remain order 1. Indeed, $\mu$ is supposed to be the relevant scale for new physics, and a large $\delta$ would make this identification suspect.

Proceeding as in \cite{Bertoldi:2010ca}, we assume that the occupation number of excitations above the ground state is $e^{-\beta \omega_n}\mathcal{F}'\left(e^{-\beta \omega_n}\right)$.  This is merely a convenience since the function $\mathcal{F}$ remains unspecified.  The only content is that the occupation number is some function of the ratio of the temperature to the energy of the state.  An important physical assumption has gone in at this point.  Essentially we have a bath of particles, with some number density $n$.  We are assuming very small temperatures so that this density may be considered very large, i.e. even though we are exciting a certain number of them above their ground state, the ``same number'' remain unexcited.  Therefore, we are treating this number density of particles as an inexhaustible bath.  This essentially means that the number of possible excitations is approximately infinite, and there is no effective chemical potential for adding new ones.  The thermodynamics is governed by some collective mode, the ``phonons'', rather than the constituents of the material, the ``molecules.''

Taking the above functional form of the occupation numbers, we may write
\be
E=-\frac{\pa}{\pa \beta} F, \qquad F=\sum_n \mathcal{F}\left(e^{-\beta \omega_n}\right).
\ee
Calculating $F$, and approximating the sum as an integral, we find
\bea
F&=&\Omega_{d-1}\int n^{d} \frac{dn}{n}{\mathcal{F}}\left(e^{-\beta\mu \frac{\delta}{(\ell \mu)^z} n^z}\right) \nn \\
&=& \frac{\Omega_{d-1}}{z} \frac{(\ell \mu)^d}{(\delta \beta \mu)^{{d}/{z}}} \int \hat{n}^{{d}/{z}} \frac{d\hat{n}}{\hat{n}}{\mathcal{F}}\left(e^{-\hat{n}}\right)
\eea
and define, therefore,
\be
\hat{F}\left(\frac{d}{z}\right)=\int \hat{n}^{{d}/{z}} \frac{d\hat{n}}{\hat{n}}{\mathcal{F}}\left(e^{-\hat{n}}\right).
\ee
Finally, we read that the energy is
\be
E=\Omega_{d-1}\frac{d}{z^2} \frac{(\ell \mu)^d}{(\delta \mu)^{{d}/{z}}}
T^{1+d/z}\hat{F}\left(\frac{d}{z}\right).
\ee
Note that the constant $\ell^d={V_d}/{(2\pi)^d}$.  Hence, we may take derivatives of the above by $T$ at constant volume in a straightforward way.  Therefore
\be
S=\int_V dT \frac{1}{T}\left(\frac{\pa E}{\pa T}\right)_V=\Omega_{d-1}\frac{d+z}{z^2} \frac{(\ell \mu)^d}{(\delta \mu)^{{d}/{z}}}\hat{F}
\left(\frac{d}{z}\right)T^{{d}/{z}}.
\ee
Happily, this exhibits the $T^{d/z}$ behaviour that we see in the $T\ll\mu$ limit of the gravity system.
Using the above expressions, one can show the expected thermodynamic relation
\be
{\mathcal E}=\frac{d}{d+z}Ts\,,
\ee
where we have divided by the volume to give the expressions in terms of the energy density ${\mathcal E}$ and entropy density $s$. The above energy that we have arrived at is the ``dynamical'' energy: it cannot be directly compared with the energy in AdS.  The absolute energy in AdS will have a constant offset, due to the finite charge density and chemical potential of the background.  Since this is some fixed ``zero point'' energy for low temperatures, it does not come into the above considerations in the effective field theory\footnote{This is akin to ignoring the rest mass 
of a non-relativistic material when computing the energy.}.  For a direct comparison to the numerical setup we would need to subtract off the asymptotic contribution: we do not do this here, and content ourselves with seeing the correct dependence of $s$ on $T$ for $T\ll\mu$, which then implies the above expression up to an additive constant.

Now let us recall that we have done the above for one type of particle.  We must multiply this by the number of species to make a meaningful comparison to the gravity setup.  In fact, the measure of the number of species is given by something like $N_{S} \sim {L^d}/{(4 G_{d+2})}$ for the gravity setup so the total entropy above gets multiplied by this amount.  Finally, dividing by volume $(2\pi \ell)^d$, we find that the total entropy density should be
\be
4 G_{d+2} s=4 G_{d+2} \frac{S}{(2\pi \ell)^d}=
\delta(z,d)^{-{d}/{z}}
\kappa(z,d)\frac{\Omega_{d-1}}{(2\pi)^d}\frac{d+z}{z^2} (L\mu)^d \hat{F}\left(\frac{d}{z}\right)\left(\frac{T}{\mu}\right)^{{d}/{z}}\,,
\ee
where on the right hand side the function $\kappa(z,d)$ measures the number of species exactly: $N_{S}=\kappa(z,d){L^d}/{(4 G_{d+2})}$.

One possible function of interest is a simple exponential suppression, ${\mathcal{F}}(x)=x$, i.e. a Boltzmann distribution.  This is in some sense a ``dilute phonon'' limit, where the excitation modes do not interact.  In such a situation we find
\be\label{eq:s-ft}
4 G_{d+2}s=\delta(z,d)^{-{d}/{z}} \kappa(z,d)\frac{\Omega_{d-1}}{(2\pi)^d}\frac{d+z}{z^2} (L\mu)^d \Gamma\left(\frac{d}{z}\right)\left(\frac{T}{\mu}\right)^{{d}/{z}},
\ee
and the above is a good unitless measure of the entropy density.  We can, of course, combine the two functions $\delta$ and $\kappa$ together into one function that measures the number of degrees of freedom.

We are now in a position to make a qualitative comparison to the graphs in Figure \ref{fig:int}.  First, we assume that $\kappa(z,d)$ and $\delta(z,d)$ are always order 1 numbers.  Hence, in the large $z$ limit, these must asymptote to some fixed value, depending only on $d$.  Therefore, all that is left to analyze is the behaviour of the factor  $\Gamma\left(\frac{d}{z}\right) (d+z)/z^2$.  In the large $z$ limit  $\Gamma\left(\frac{d}{z}\right) (d+z)/z^2\rightarrow \frac{1}{d}$ and so we see that the whole expression (\ref{eq:s-ft}) goes to a $z$ independent value, depending only on $d$.  Therefore, the graphs in Figure \ref{fig:int} qualitatively agree with the large $z$ behaviour of equation (\ref{eq:s-ft}), given the restriction on $\delta(z,d)$ and $\kappa(z,d)$.

\section*{Acknowledgements}

This work was funded by a grant from NSERC of Canada. The work of IGZ was also supported by an Ontario Graduate Scholarship.

\end{document}